Transport in gapped bilayer graphene: the role of potential fluctuations

Ke Zou and Jun Zhu

Department of Physics, The Pennsylvania State University, University Park, PA 16802-6300

**Online Supporting Information**

1. The energy and size distribution of electron and hole puddles in bilayer graphene.

2. The calculation of $<\varepsilon_{ij}>$.

3. Raman spectra on dual-gated bilayer graphene.

## 1. The energy and size distribution of electron and hole puddles in bilayer graphene:

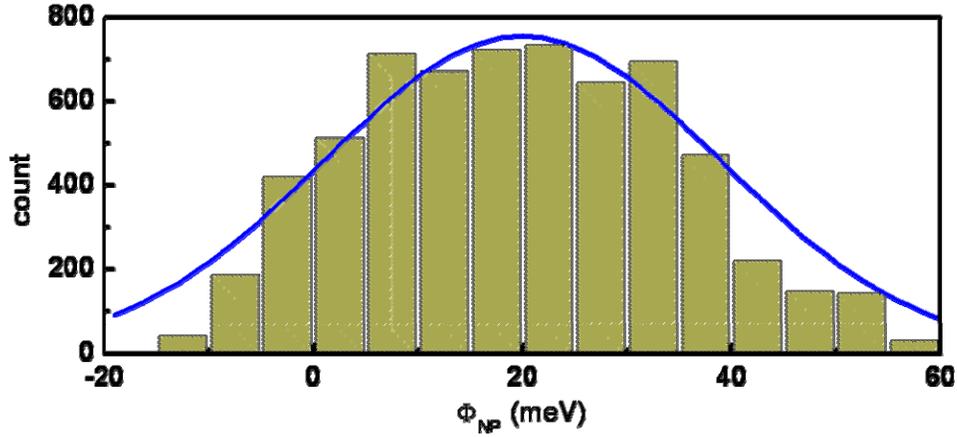

**Figure S1. The histogram of $\Phi_{NP}$ in Fig. 3 of Ref. 1. The blue solid line is a Gaussian fit.**

Figure S1 plots a histogram of the charge neutrality point (CNP) potential $\Phi_{NP}$ of bilayer graphene on $SiO_2$, extracted from scanning tunneling spectroscopy data reported in Fig. 3 of Ref. 1. The solid line is a Gaussian fit which yields $\Phi_{NP} = (20\pm19)$ meV.

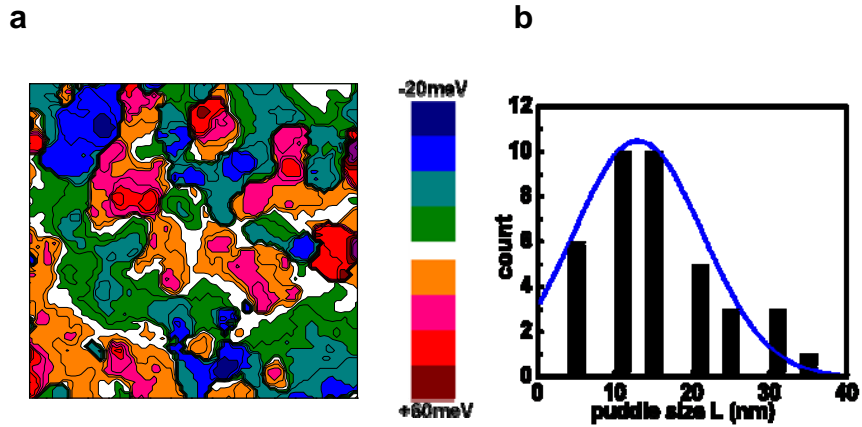

**Figure S2: (a) $\Phi_{NP}$ data in Fig. 3 of Ref. 1. The map is 80 x 80nm. Each color represents a 10 meV step. (b) The histogram of puddle size *L* from (a). The blue solid line is the best fit by a Gaussian distribution and yields *L* = (13 ± 8.4) nm.**

Figure S2 (a) shows the spatial map of the same $\Phi_{NP}$ data in a false color plot. The white area corresponds to 20±2.5 meV, representing a small gap separating electron and hole puddles. We approximate each puddle as a polygon and estimate the lateral size *L*. Puddles of the same carrier type separated by a barrier of less than 5 meV are joined together and counted as one large puddle. The histogram of *L* obtained from both electron and hole puddles and from both dimensions is shown as Fig. S2 (b). A truncated and normalized Gaussian fit yields a distribution of

$$P(L) = \frac{1.06}{\sqrt{2\pi\sigma^2}} \exp(\frac{-(L-L_0)^2}{2\sigma^2}), \qquad (1)$$

where $L_0 = 13$ nm and $\sigma = 8.4$ nm for $L > 0$ and $P(L)=0$ for $L \leq 0$.

## 2. The calculation of $<\varepsilon_{ij}>$:

We simplify the fluctuating potential as a network of cylindrical potential wells with a fixed depth $V_0$ but varying size *L* (Fig. 5(a) in the text and Fig. S3) and solve for the eigenenergies of the bound states for different *L*. The resulting $\varepsilon_i (L)$ for $V_0 = 13$ meV and 43 meV are shown in Fig. 5 (b) of the text.

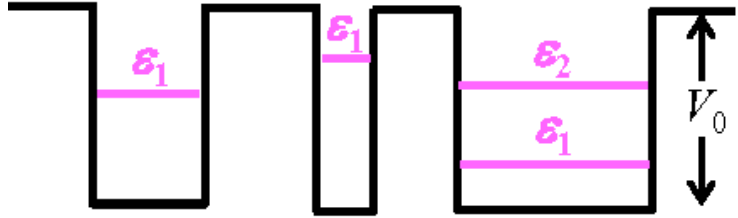

**Figure S3. Localized states in cylindrical potential wells. A narrower well has a higher ground state energy and a wider well may have more than one bound state.**

The average energy difference between neighboring states $<\varepsilon_{ij}>$ is calculated using the following equation:

$$<\varepsilon_{ij}> = <|\varepsilon_i(L_i) - \varepsilon_j(L_j)|> = \iint_{Li,Lj} |\varepsilon_i(L_i) - \varepsilon_j(L_j)| P(L_i) P(L_j) dL_i dL_j. \qquad (2)$$

For wells with 2 bound states, the 1$^{st}$ excited state is used in the calculation.

## 3. Raman spectra on dual-gated bilayer graphene.

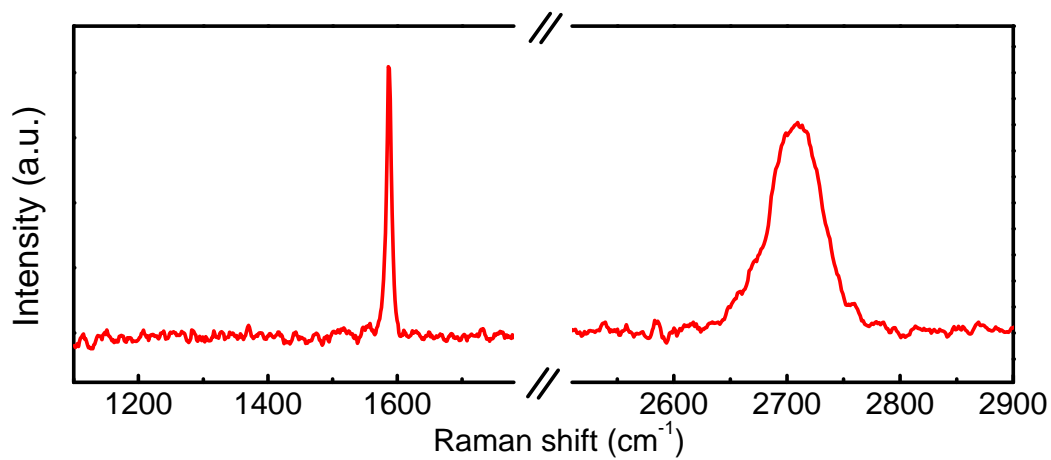

**Figure S4.** Raman spectra on sample A1 taken with 514 nm laser excitation. The horizontal axis scale is different for the G (1580 cm$^{-1}$) and 2D (~ 2700 cm$^{-1}$) regions.

Figure S4 shows a Raman spectrum taken on sample A1. The spectrum shows the characteristic 2D band of bilayer graphene. No visible D band (1360cm$^{-1}$) is observed.

[1] A. Deshpande, W. Bao, Z. Zhao, C. N. Lau, and B. J. LeRoy, Appl. Phys. Lett. **95**, 243502 (2009).